\documentclass{aa}  
\usepackage{natbib,twoopt}
\bibpunct{(}{)}{;}{a}{}{,} 

\makeatletter
\newcommandtwoopt{\citeads}[3][][]{\href{http://adsabs.harvard.edu/abs/#3}%
{\def\hyper@linkstart##1##2{}%
\let\hyper@linkend\@empty\citealp[#1][#2]{#3}}}
\newcommandtwoopt{\citepads}[3][][]{\href{http://adsabs.harvard.edu/abs/#3}%
{\def\hyper@linkstart##1##2{}%
\let\hyper@linkend\@empty\citep[#1][#2]{#3}}}
\newcommandtwoopt{\citetads}[3][][]{\href{http://adsabs.harvard.edu/abs/#3}%
{\def\hyper@linkstart##1##2{}%
\let\hyper@linkend\@empty\citet[#1][#2]{#3}}}
\newcommandtwoopt{\citeyearads}[3][][]%
{\href{http://adsabs.harvard.edu/abs/#3}
{\def\hyper@linkstart##1##2{}%
\let\hyper@linkend\@empty\citeyear[#1][#2]{#3}}}
\makeatother

\usepackage{graphicx}
\usepackage{color}
\usepackage[varg]{txfonts}
\usepackage{hyperref}   

\usepackage{arydshln}
\usepackage{multirow}

\usepackage{float} 

\begin{document} 

   \title{VarIabiLity seLection of AstrophysIcal sources iN PTF (VILLAIN) \\ 
   II. Supervised classification of variable sources 
}

   \author{
           S. H. Bruun\inst{1}
           \and J. Hjorth\inst{1}
           \and
           A. Agnello\inst{1,2}
           }

   \institute{DARK, Niels Bohr Institute, University of Copenhagen,
              Jagtvej 128, 2200 Copenhagen N, Denmark
              \and
              STFC Hartree Centre, Sci-Tech Daresbury, Keckwick Lane, Daresbury, Warrington (UK) WA4 4AD
              \\
              \email{ sofie.bruun@nbi.ku.dk, adriano.agnello@stfc.ac.uk, jens@nbi.ku.dk}
              }

\date{April 19 2023}

    \titlerunning{Variability selection in PTF: II. Classification of lightcurves}

  \abstract
   { Large, high-dimensional astronomical surveys require efficient data analysis. Automatic fitting of lightcurve variability and machine learning may assist in identification of sources including candidate quasars.  }
   {We aim to classify sources from the Palomar Transient Factory (PTF) as quasars, stars or galaxies, and to examine model performance using variability and colours. We determine the added value of variability information as well as quantifying the performance when colours are not available.}
   {We use supervised learning in the form of a histogram-based gradient boosting classifier to predict spectroscopic SDSS classes using photometry. For comparison, we create models with structure function variability parameters only, magnitudes only and using all parameters.}
   { We achieve highly accurate predictions for 71 million sources with lightcurves in PTF. The full model correctly identifies 92.49~\% of spectroscopically confirmed quasars from the SDSS with a purity of 95.64~\%. With only variability, the completeness is 34.97~\% and the purity is 58.71~\% for quasars. The predictions and probabilities of PTF objects belonging to each class are made available in a catalogue, VILLAIN-Cat, including magnitudes and variability parameters.}
   { We have developed a method for automatic and effective classification of PTF sources using magnitudes and variability. For similar supervised models, we recommend using at least 100\,000 labeled objects, and we show how performance scales with data volume.  } 


   \maketitle

\section{Introduction}

Machine learning has gained increasing importance in astronomy \citep{NN_astro,astroML2010,astroML2022}. In the era of large astronomical surveys, automatic classification is necessary for fast and reliable processing of sources, and for detecting patterns in high-dimensional data. Detailed observations including spectroscopy are expensive for large datasets, so high-quality photometric classification is needed for future surveys such as the Vera C. Rubin Observatory Legacy Survey of Space and Time \citep[LSST;][]{LSST}. 

Quasars and active galactic nuclei (AGN) 
can be identified based on their variability. Variability is especially important for classification of AGN in low mass galaxies, as they are often missed using emission line ratios. This is due to low metallicities (weakening the [\ion{N}{II}] line), different AGN fuelling mechanisms or "star formation dilution" -- dilution of the optical emission line signature of AGN by \ion{H}{II} region emission lines in galaxies with high star formation rates \citep{lowmassAGN,SFdilution,lowmetalAGN}. \citet{TESSvarAGN} selected candidate AGN using optical variability of 142\,061 lightcurves from the Transiting Exoplanet Survey Satellite \citep{TESS} using parameter cuts and visual inspections. Out of the 29 AGN candidates, 8 are in low mass galaxies, but the method would be impractical for larger datasets. 

Various machine learning models have been applied to photometric data 
in search of quasars and AGN candidates. \citet{cicco2021} selected optically variable AGN candidates with a Random Forest algorithm \citep{RF} on measures that will available with the LSST: lightcurves and colours in the optical and near-infrared. 
\citet{palanque} used neural networks 
to distinguish high-redshift quasars and stars in Stripe 82 from the Sloan Digital Sky Survey \citep[SDSS;][]{stripe82} and in simulated Palomar Transient Factory \citep[PTF;][]{PTF,Rau} data. Incorporating variability parameters improved the selection purity for quasars at high redshifts and quasars with broad absorption lines.

\citet{SHEEP} applied gradient boosting algorithms for SDSS class prediction on SDSS and WISE photometry. The model performed much better using combinations of magnitudes to create colours and photo-z predictions than without the combinations. The photo-z model was trained to predict spectroscopic redshifts from SDSS.

In this paper, we aim to create a large catalogue of candidate quasars, stars and galaxies, and to analyse the importance of monochromatic lightcurve variability compared to the importance of optical and infrared magnitudes and colours using a machine learning model.  We use the full set of lightcurves from PTF fitted with power-law variability models by \citet{paper1}; paper I of the VILLAIN project.

In Sect. \ref{sec:method} we define the machine learning models, how they are selected and the training strategy. The section includes definitions of model inputs, preprocessing to create colours from magnitudes etc., and we define the metrics for model evaluation. An overview of the results including model performances is presented in Sect. \ref{sec:results}. In Sect. \ref{sec:discussion}, we discuss the results, biases and perspectives for the use of the model predictions and adaptions of the method for different contexts.

\section{Method} \label{sec:method}

To define a model that can classify objects, we need a set of objects to learn from. In the present paper, we assume spectroscopically confirmed classifications in SDSS to be the ground truth. Each object is classified by SDSS as either a quasar, star or galaxy, which are the "labels" of the model. To learn how to reproduce the labels without spectroscopy, we build a model to guess the labels using data from PTF, Wide-field Infrared Survey Explorer \citep[WISE;][]{WISE} and Pan-STARRS1 \citep[PS1;][]{PS1survey}. We use a set of 70\,920\,904 PTF sources with power law fits of their structure functions. The PTF sources have been cross-matched to sources in the WISE, PS1 and SDSS, as described by paper I.

We refer to each object in the dataset -- and all parameters associated with it -- as a "sample", and each sample corresponds to a PTF object. A sample is described by a number of properties called "features" which are used as inputs for the machine learning model, e.g. magnitudes.
We can split the set of samples into a labeled set and an unlabeled set. 2.5~\% of PTF sources have matches in SDSS PhotoObj and SpecObj, resulting in a labeled set of 1\,747\,471 sources. 

A supervised machine learning algorithm takes input features $X$ and predicts labels $Y_{pred}$ after learning on "true" labels $Y_{true}$. The input "true" labels are spectroscopic classifications from SDSS. The model only learns from the labeled subset, but is able to predict labels for every source. While fitting (also called "training") the model to data, model performance is quantified by a loss function, which measures dissimilarity between the true labels and predicted labels. The loss is minimised to improve classification.

It is possible to learn from the unlabled subset as well. An unsupervised machine learning model would learn patterns in $X$ and could group sources with similar parameters into "clusters". To learn from both unlabeled data and the labels, when they are available, semisupervised learning is preferable. One form of semisupervised learning is self training which iteratively learns from the labeled samples and assigns probable pseudo-labels to unlabeled samples. Due to the large dataset of this project, self training did not improve performance (F1 scores as described in Sect. \ref{sec:performance}). We therefore proceed with supervised learning.

\subsection{Models} \label{sec:models}
Machine learning models learn parameters from the input data, but they also have hyperparameters to control the learning process. To choose an algorithm and its hyperparameters, testing multiple models is beneficial. 
To compare the performance of models with and without variability and magnitude information, we also create multiple final models for different datasets using different features. 
The datasets include samples with missing values in WISE and PS1, so for fair evaluation of the importance of colours, we create different models for the full dataset and for the "matched" dataset. The six models use:
\begin{enumerate}
    \item AllA: \textbf{All} features and \textbf{A}ll data (only PTF lightcurve variability and SDSS matches are required)
    \item AllM: \textbf{All} features and only data with \textbf{M}atches in WISE and PS1 (so their magnitudes are always included)
    \item VarA: Only \textbf{Var}iability features and \textbf{A}ll data
    \item VarM: Only \textbf{Var}iability features and only data with \textbf{M}atches in WISE and PS1
    \item MagA: Only \textbf{Mag}nitude features and \textbf{A}ll data
    \item MagM: Only \textbf{Mag}nitude features and only data with \textbf{M}atches in WISE and PS1
\end{enumerate}
We create additional models for AllA using random subsets of varying size to analyse the importance of data volume.

\subsection{Data splitting}
The goal is to create a model that can reliably classify data it has not seen before. It should give predictions based on general trends in the data and avoid overfitting to random patterns in a specific dataset even if it can perfectly replicate SDSS labels in data it was trained on. To get an unbiased measure of model performance, we set aside part of the labeled data as a test set for evaluation. The model is only tested on the test set once, and we do not examine its properties prior to this test, except size and number of matches in WISE and PS1, to avoid any leakage of information from the test set into the model. 

During model optimisation, we train the model multiple times and determine when it is ready to run on the test set. We evaluate performance on the training set, but this is prone to overfitting. For a less biased estimate, we use a validation set. While trying different models and tuning hyperparameters, we evaluate on the validation set. Generally, the performance will seem best on the training set and worst on the test set. Using the test set multiple times would likely give seemingly better scores as well, but it would be a biased estimate of performance on new data. For further discussion of best practices in machine learning, see \citet{MLguide} and \citet{MLguide2}.

Before testing, we could combine the training and validation sets to train one last time on more data. We choose not to combine them, because the new model randomly could be worse (and we would not know without a separate validation set) and because some models are sensitive to the size of the dataset. For small datasets, the advantage of learning from more data would outweigh this concern. 

We shuffle the data and use a 60-20-20 split for creating a training set, a validation set and a test set. This is done both for the full dataset and the matched subset. We make sure that the training set for the matched subset, is also a subset of the training set for the full dataset, and likewise for validation and test sets.

\subsubsection{Cross-validation}

An alternative splitting technique for training and validation is cross-validation \citep{CV}, which allows the use of all data for both purposes. In $n$-fold cross validation, the data is split into $n$ folds. Alternately, $n-1$ folds are used for training and one for validation. The average score of the validation folds is then used for evaluation -- a process which reduces variance compared to a single validation set.
We use cross-validation within the training set for less overfitting, but still keep extra validation and test sets, since we run the cross-validation multiple times for adjusting hyperparameters. 
The balance of class frequencies (fraction of samples of each class) in the dataset will affect the model. 
During cross-validation, we ensure all folds are representative of the full dataset by using stratified folds, meaning that the class balance of the full set is preserved in each fold.

\subsubsection{Early stopping}

Reducing overfitting on the training set will likely lead to better performance on the test set, because when the model mistakenly thinks it perfectly understands the training data, it will stop learning. We use several methods to avoid overfitting. One of them involves data splitting, namely early stopping. 

To stop the algorithm before it overfits, we take out part of the training set as a validation set for evaluating performance after every step. If performance does not improve by more than a predefined tolerance, training is stopped. We implement the early stopping and cross-validation in Sect. \ref{sec:hyperpar}. 

\subsection{Classification algorithm} \label{sec:algorithm}

Astronomical datasets often include objects with missing measurements of some properties. 
To analyse a dataset with missing feature values, one could drop objects with missing features or drop features that are not available for all objects. Another option is to guess the missing features (imputation). To learn from all data without dropping or guessing information, we use a model that by construction accepts and learns from missing values \citep{josse,twala}.  

For fast classification with low memory usage and including non-linear patterns, we choose \texttt{HistGradientBoostingClassifier}, which is a tree based model from scikit-learn \citep{sklearn}. Tree based models classify data using hierarchical tree-structures. A decision tree 
places "nodes" which split the feature space, and the nodes are used one by one to decide which class a sample belongs to. The final nodes, which are not split further, are called leaf nodes, while the internal nodes are called branch nodes. Sources are classified based on which leaf they belong to. 

A single decision tree is simple and easy to interpret, however, ensemble tree models will usually perform better. 
One way to improve the outputs of a tree $f_i$, is to add another tree $h_{i+1}$ that predicts how the first one could be improved, and create a new ensemble model $f_{i+1}$. Gradient Boosting Machines sequentially add trees in this manner. The loss function $L$ is minimised to improve classification, using the gradients $G = \nabla L(Y,f(X))$ and hessians $H = \nabla^2 L(Y,f(X))$ of $L$ with respect to the model $f(X)$. 
First, constant initial predictions $C$ are chosen. 
Then, a tree $h_1(X)$ fits the gradients of the constant model $f_0=C$, thereby predicting a correction that can be added to the initial predictions. The next tree $h_{2}$ predicts gradients of the updated model $f_{1}$ and so forth. This continues for $max\_iter$ boosting iterations with the final predictions $F(x)$ including all correcting trees multiplied by the learning rate, $\eta$ \citep{learningrate}:
\begin{align}
    F(X) = C + \eta \sum^{max\_iter}_i h_i (X).
\end{align}
The learning rate is a regularisation parameter that prevents overfitting by preventing the model from learning too much from a single tree. 

\subsubsection{Histogram-based Gradient Boosting Classification Tree} \label{sec:HGBC}

\texttt{HistGradientBoostingClassifier} is a scikit-learn implementation of gradient boosting similar to LightGBM \citep{LightGBM}. In this tree-based ensemble method, each sample is processed by a series of trees. In each tree, the sample is assigned to a leaf which has a leaf weight. The leaf weights can then be summed for prediction of the class of the sample. To convert the sum to a class for multi-class prediction, a tree is created for each class during each iteration. The values can then be compared following a one versus rest (OvR) approach in which every class is compared against all other classes. Softmax normalisation converts the values to probabilities.

The trees are estimating corrections to previous trees, so the leaf weights depend on gradients and hessians of the loss function of a tree $i$:
\begin{align}
    w_i = -\frac{G_i}{H_i+\lambda}.
\end{align}
Here $\lambda$ is a regularisation parameter used for penalising complex models to avoid overfitting, similarly to $\eta$. It is also part of the loss function. For multiclass classification, we define the loss function $L$ as the categorical crossentropy between a tree model $f$ and "true" labels $Y$ plus a regularisation term $\Omega$ for each tree $h_j$. $L$ is computed for $N$ samples as:
\begin{align}
    L(Y,f(X)) &= - \frac{1}{N}\sum_i\sum_k Y_{i,k} \ln(f_{i,k}(X_i)) + \sum_j \Omega(h_j), \label{eq:crossent} \\
    \Omega(h_j) &= 
    \frac{1}{2}\lambda ||w_j||^2, \label{eq:regularisation}
\end{align}
where $f_{i,k}$ is the predicted probability for a sample $i$ of belonging to class $k$, and $Y_{i,k}$ is the "true" one-hot-encoded value (0 or 1) we are trying to predict. For regularisation, $\lambda$ penalises trees with high $||w_j||^2$ to avoid large contributions from individual trees \citep{XGBoost}. 

When the model decides where to split the data (creating nodes) for each tree, \texttt{HistGradientBoostingClassifier} speeds up the process by binning the feature space instead of sorting all feature values. The algorithm creates histograms of $G$ and $H$ for all samples in the feature bins, and uses the histograms to decide at which bin edge to split. 
A split optimises the gain of the left and right nodes,
\begin{align}
    \text{Gain} = \frac{1}{2} \left( \frac{G_{\text{left}}^2}{H_{\text{left}} + \lambda} + \frac{G_{\text{right}}^2}{H_{\text{right}} + \lambda} - \frac{(G_{\text{left}} + G_{\text{right}})^2}{H_{\text{left}} + H_{\text{right}} + \lambda} \right).
\end{align}
To include samples with missing values, one bin is dedicated to this, which is a "missingness incorporated in attributes" strategy \citep{twala}. At every split, the model learns which nodes to assign sources with missing values to. This strategy takes into account the information on whether data is missing or not, and allows a model trained on all features of Sect. \ref{sec:preproc} to classify sources even if they have no colour information. 

\subsubsection{Hyperparameter tuning} \label{sec:hyperpar}

For fast hyperparameter tuning, we use successive halving iterations \citep{halving1,halving2}. The \texttt{HalvingRandomSearchCV} class in scikit-learn randomly chooses hyperparameter values in specified ranges and examines performances using 5-fold cross-validation. The model is evaluated with a low number of samples at first. Then the best performing hyperparameters are selected, and they are evaluated with a higher number of samples until the best hyperparameter combination is determined. Performance in HalvingRandomSearchCV() is evaluated with the macro averaged F1 score as defined in Sect. \ref{sec:performance}.

We use \texttt{HistGradientBoostingClassifier} with early stopping, which sets aside 10~\% of the data for validation, and stops the fitting if the last 30 models did not improve the loss function by more than $10^{-7}$.
With $max\_iter$ at 200, we test learning rates from 0.01 to 0.3, maximum number of leaf nodes from 11 to 81, minimum samples per leaf from 5 to 200 and l2 regularisation parameters ($\lambda$ in Eq. \ref{eq:regularisation}) from 0 to 5.

\subsection{Preprocessing} \label{sec:preproc}

The SDSS labels are quasar, star and galaxy, and we choose to encode them as 0, 1 and 2, respectively, in the $Y_{true}$ vector. The model converts $Y_{true}$ to $Y_{i,k}$ of Eq. \ref{eq:crossent}. 
We choose input features that are suitable for class prediction by being measured independently of the spectroscopic classes (unlike e.g. spectroscopic redshift).
We include variability parameters $A$ and $\gamma$, $W1$ and $W2$ from WISE, $g, r, i, z$ and $y$ from PS1 and median $R$ from PTF. 
In tree based models, scaling the features would also scale the positions of tree nodes in feature space. This does not affect performance, so the models are insensitive to monotonic feature transformations, and the features will not need scaling.

\subsubsection{Feature engineering} \label{sec:features}

Features may interact in ways that are useful for classification. To make it easier for a model to learn such interactions, we construct additional features by combining existing ones. We create colour features and features based on the selection criteria of paper I and \citet{schmidt}.

In paper I, manually chosen selection criteria are effective in selecting pure sets of each class. The criteria are applied in $\log (A)$ vs. $\gamma$, $z-W1$ vs. $g-r$ and $W2$ vs. $W1-W2$. Here $\log (A)$ is in base 10. The criteria are linear in said parameter spaces. For easier separation of classes, we construct four features  
$z-W1-1.25 (g-r), W1-W2-0.017W2, \gamma+0.5\log (A)$ and $\gamma-2\log (A)$. 
We also combine all eight magnitude features for 28 colour features and thereby use a total of 42 features. For non-linear feature construction, the features could be combined by multiplication as well, but this did not give an improvement compared to models with only linear constructed features. 

\subsection{Performance evaluation} \label{sec:pereval}
To measure multiple qualities of the predicted outputs, the model performance is evaluated with multiple metrics. 
Mean accuracy (the fraction of correct predictions) is simple, but highly dependent on class ratios.
The macro-averaged scores are the unweighted means of the scores for each class and assume equal importance of each class.

\subsubsection{F1 score}

F1 is the harmonic mean of completeness and purity \citep{Dice, sorensen}, which are assumed equally important. They are computed from the number of true positives (TP), false negatives (FN), and false positives (FN):
\begin{align}
    \mathrm{Completeness} &= \frac{\mathrm{TP}}{\mathrm{TP+FN}}, \\
    \mathrm{Purity}       &= \frac{\mathrm{TP}}{\mathrm{TP}+\mathrm{FP}},\\
    \mathrm{F1}       &= 2\frac{\mathrm{Completeness}\times\mathrm{Purity}}{\mathrm{Completeness + Purity}}.  \label{eq:f1}
\end{align}

\subsubsection{ROC AUC} \label{sec:roc}
ROC AUC is the Area Under the Curve (AUC) for the Receiver Operating Characteristic \citep[ROC;][]{ROC}. A ROC curve evaluates the trade-off between a high true positive rate (TPR, completeness) and low false positive rate for a binary classifier. The false positive rate, FPR, is
\begin{align}
    \mathrm{FPR} &= \frac{\mathrm{FP}}{\mathrm{FP+TN}},
\end{align}
where TN are the true negatives.
The classifier of this project is multinomial, so we compute the ROC AUC for each class with an OvR approach. 

For the average ROC AUC, to avoid the influence of class imbalances, the macro average is taken following a one versus one approach \citep[OvO;][]{handtill}:
\begin{align}
    M = \frac{2}{c (c-1)}\sum_{i<j}\hat{A}(i,j),
\end{align}
where $c$ is the number of classes and $\hat{A}$ is a measure on $i$ and $j$ -- in this case $\hat{A}$ is ROC AUC. This takes into account that each class can have better separation from one class than the other.
  
\begin{table*}[]
      \centering
      \caption{Classifier hyperparameters.}
      \begin{tabular}{l l l l l l l}
          \hline\hline   
          Model features & \multicolumn{2}{c}{Variability} & \multicolumn{2}{c}{Colours} & \multicolumn{2}{c}{All features}\\
          Samples & All  & Matched  & All & Matched  & All & Matched \\ 
          \hline
          $\lambda$            & 0.918 & 2.38 & 3.06 & 2.41 & 4.03 & 2.63 \\
          $\eta$               & 0.0462 & 0.0215 & 0.0498 & 0.0415 & 0.142 & 0.0653 \\
          Max leaf nodes       & 11 & 23 & 45 & 63 & 58 & 61\\
          Min samples per leaf & 41 & 37 & 7 & 19 & 6 & 19
      \end{tabular}
      \label{tab:hyperpar}
      \tablefoot{Hyperparameters of the histogram-based gradient boosting classifier are selected as described in Sect. \ref{sec:algorithm}. Here "colours" include all magnitudes and colours.
      }
  \end{table*}{}

\section{Results} \label{sec:results}

We tune the models in Sect. \ref{sec:models} to the training set and validate on the validation set with the chosen hyperparameters listed in Table \ref{tab:hyperpar}.
Afterwards, we run them once on the test set for each final model. This is to test performance, classify all samples and predict probabilities of being a quasar, star or galaxy. 

\begin{table*}[]
      \centering
      \caption{Performance statistics on the test set.}
      \begin{tabular}{l l l l l l l l l }
          \hline\hline
          Model features & \multicolumn{2}{c}{Variability} & \multicolumn{2}{c}{Colours} & \multicolumn{2}{c}{All features} & \multicolumn{2}{c}{Uniform baseline} \\
          Samples               & All & Matched & All & Matched & All & Matched & All & Matched\\ 
          \hline
          ROC AUC  \\
          \hspace{0.5cm}  OvO macro avg & 0.7671 & 0.7995 & 0.9945 & 0.9962 & 0.9945 & 0.9962 & 0.50 & 0.50\\
          \hspace{0.5cm}  Quasars & 0.8182 & 0.8251 & 0.9932 & 0.9951 & 0.9931 & 0.9951 & 0.50 & 0.50\\
          \hspace{0.5cm}  Stars & 0.7274 & 0.7864 & 0.9969 & 0.9976 & 0.9969 & 0.9976 & 0.50 & 0.50\\
          \hspace{0.5cm}  Galaxies & 0.6997 & 0.7268 & 0.9954 & 0.9956 & 0.9953 & 0.9957 & 0.50 & 0.50\\
          F1 & & & & & & \\
          \hspace{0.5cm}  macro avg & 0.5431 & 0.5693 & 0.9624 & 0.9733 & 0.9639  & 0.9738 & 0.3039 & 0.2966 \\
          \hspace{0.5cm}  Quasars & 0.4383 & 0.4462 & 0.9379 & 0.9570 & 0.9404 & 0.9580 & 0.2094 & 0.2086 \\
          \hspace{0.5cm}  Stars & 0.4195 & 0.4596 & 0.9642 & 0.9755 & 0.9656 & 0.9758 & 0.2679 & 0.2361 \\
          \hspace{0.5cm}  Galaxies & 0.7714 & 0.8022 & 0.9853 & 0.9875 & 0.9857 & 0.9878 & 0.4345 & 0.4423 \\
          Completeness \\
          \hspace{0.5cm}  Quasars & 0.3497 & 0.3515 & 0.9220 & 0.9505 & 0.9249 & 0.9511 & 0.33 & 0.33\\
          \hspace{0.5cm}  Stars & 0.3308 & 0.3864 & 0.9702 & 0.9722 & 0.9716 & 0.9725 & 0.33 & 0.34\\
          \hspace{0.5cm}  Galaxies & 0.8682 & 0.8769 & 0.9871 & 0.9900 & 0.9875 & 0.9903 & 0.33 & 0.33\\
          Purity \\
          \hspace{0.5cm}  Quasars & 0.5871 & 0.6107 & 0.9543 & 0.9636 & 0.9564 & 0.9650 & 0.1527 & 0.1528 \\
          \hspace{0.5cm}  Stars & 0.5732 & 0.5671 & 0.9582 & 0.9788 & 0.9597 & 0.9790 & 0.2240 & 0.1822 \\
          \hspace{0.5cm}  Galaxies & 0.6940 & 0.7392 & 0.9834 & 0.9850 & 0.9839 & 0.9852 & 0.6232 & 0.6613 \\
          Mean accuracy & 0.6687 & 0.7064 & 0.9734 & 0.9806 & 0.9744 & 0.9810 & 0.33 & 0.33
      \end{tabular}
      \label{tab:scores}
      \tablefoot{Classification performance of on the test set including all samples or only those matched in WISE and PS1. Performance improves with more input features, especially "colours" including magnitudes. The baseline is described in Sect. \ref{sec:baseline}. 
      }
  \end{table*}{}

\begin{figure}[]
	\centering
    \resizebox{\hsize}{!}{\includegraphics{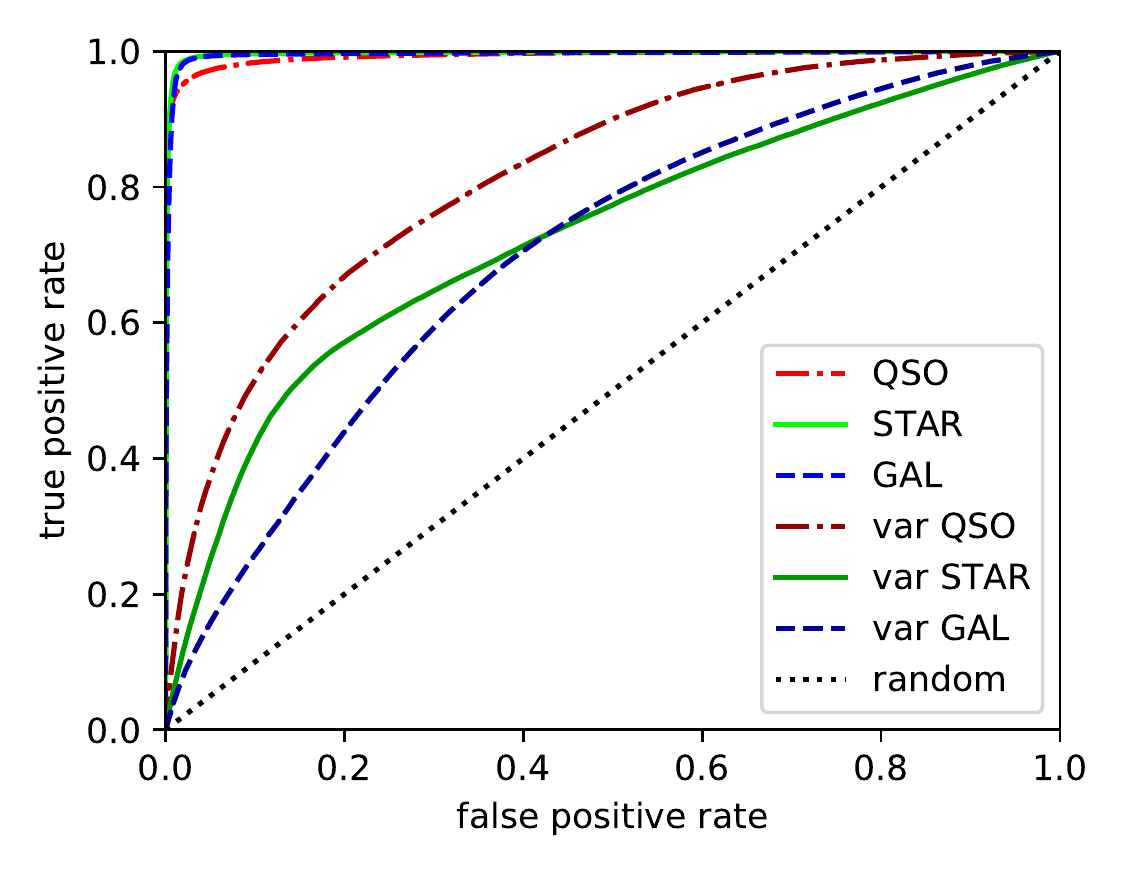}}
	\caption{ROC curves for the model with all features and the model with just variability ("var" in the legend) on all data. Including colours makes it possible to achieve both very high true positive rates and low false positive rates for both quasars (QSO), stars (STAR) and galaxies (GAL). The dotted line shows the performance of random guessing which is worse than for all trained models.} 
    \label{fig:ROC}
\end{figure}

\subsection{Performance} \label{sec:performance}

In Table \ref{tab:scores}, the models are listed along with several test statistics described in Sect. \ref{sec:pereval}. Performance is measured on the test set and is slightly lower than on the training and validation sets due to overfitting, as expected. 
The "matched" samples are matched in both PS1 and WISE, so some "unmatched" samples do have colours and magnitudes.

Fig. \ref{fig:ROC} shows test set ROC curves for each class in two models using all samples. The model using all features (AllA) shows almost perfect performance, while the model with just variability (VarA) has a greater trade off between a high TPR and low FPR. This is also reflected in the macro averaged ROC AUC scores of 0.9945 for AllA and 0.7671 for VarA. Both models are much better than random guessing, since the ROC AUC are larger than 0.5 for all classes in Table \ref{tab:scores}.

\subsubsection{Baseline} \label{sec:baseline}

In Table \ref{tab:scores} we include a baseline model that randomly assigns labels in a uniform manner. Random labels result in completenesses of 0.33 and purities equal to the frequencies of each class (but sensitive to differences between the training and test sets). The macro averaged F1 score is 0.30, which is much lower than for all trained models.

Another baseline could be labelling everything as the most frequent class: galaxies. Completeness is then 0 for quasars and stars, and 1 for galaxies. Galaxy purity is 0.62 (their frequency), and this gives an F1 score of 0.77 for galaxies. We get undefined purities and F1 scores for quasars and stars, but if we set the undefined F1-scores to 0, the macro averaged F1 score is 0.26 
using all data. 

All models perform better than the baseline models. For some metrics of some classes, VarA only matches the baselines -- but it simultaneously performs better for other classes with the same metrics.

\subsubsection{Number of samples} \label{sec:size}

\begin{figure}[]
	\centering
    \resizebox{\hsize}{!}{\includegraphics{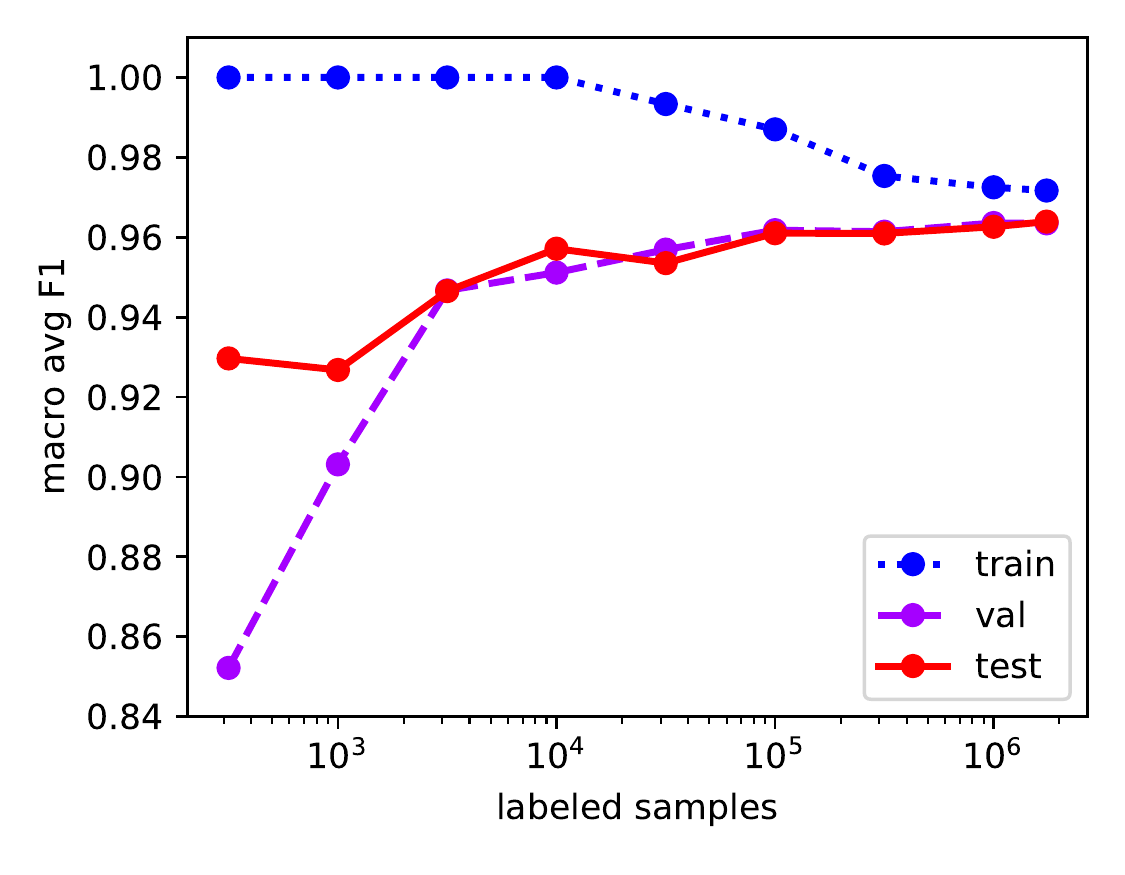}}
	\caption{The macro averaged F1 score rises with the number of labeled samples for the validation and test sets.  We see a clear increase up to $\sim $100\,000 labeled samples, and gap between performance on training and test data continues to narrow until we reach best model on the full labeled dataset of 1.7 million samples.} 
    \label{fig:N_labeled}
\end{figure}

To evaluate how many samples are needed in future surveys, we create new models using all features on random subsets of the full dataset. We estimate performance using macro averaged F1 scores. We still use a 60-20-20 split and choose random samples within the original training, validation and test sets. 

In Fig. \ref{fig:N_labeled}, we see how a larger training set leads to better performance.
For less than $10^4$ labeled samples, the training set (<6\,000 samples) is completely overfitted with a perfect F1 score of 1. For larger samples, F1 decreases for the training set and increases for the validation and test sets, except for a few cases due to randomness in which sources are included. Randomness also leads to much better scores on the test set than on the validation set for 316 and 1000 samples. For 316 samples, the main difference is that test set includes stars that are easier for the model to classify. With 1\,000 samples, the model performed better on both the galaxies and stars in the test set and it contained fewer quasars. Larger partitions for validation and test could reduce the uncertainty at the cost of less training data. 
Very similar scores on the validation and test set for the full model indicates that the amount of overfitting is similar. It indicates that the validation set could have been used more during model selection and hyperparameter tuning without becoming too biased.

The greatest improvements are for seen up to $\sim 10^5$ labeled samples. Even with just 316 labeled samples, the F1 score on the test set is 0.93. We created a model with 100 labeled samples as well, but it only predicted galaxies although the training set included all classes.

\subsection{Calibration} \label{sec:calibration}

\begin{figure}[]
	\centering
    \resizebox{\hsize}{!}{\includegraphics{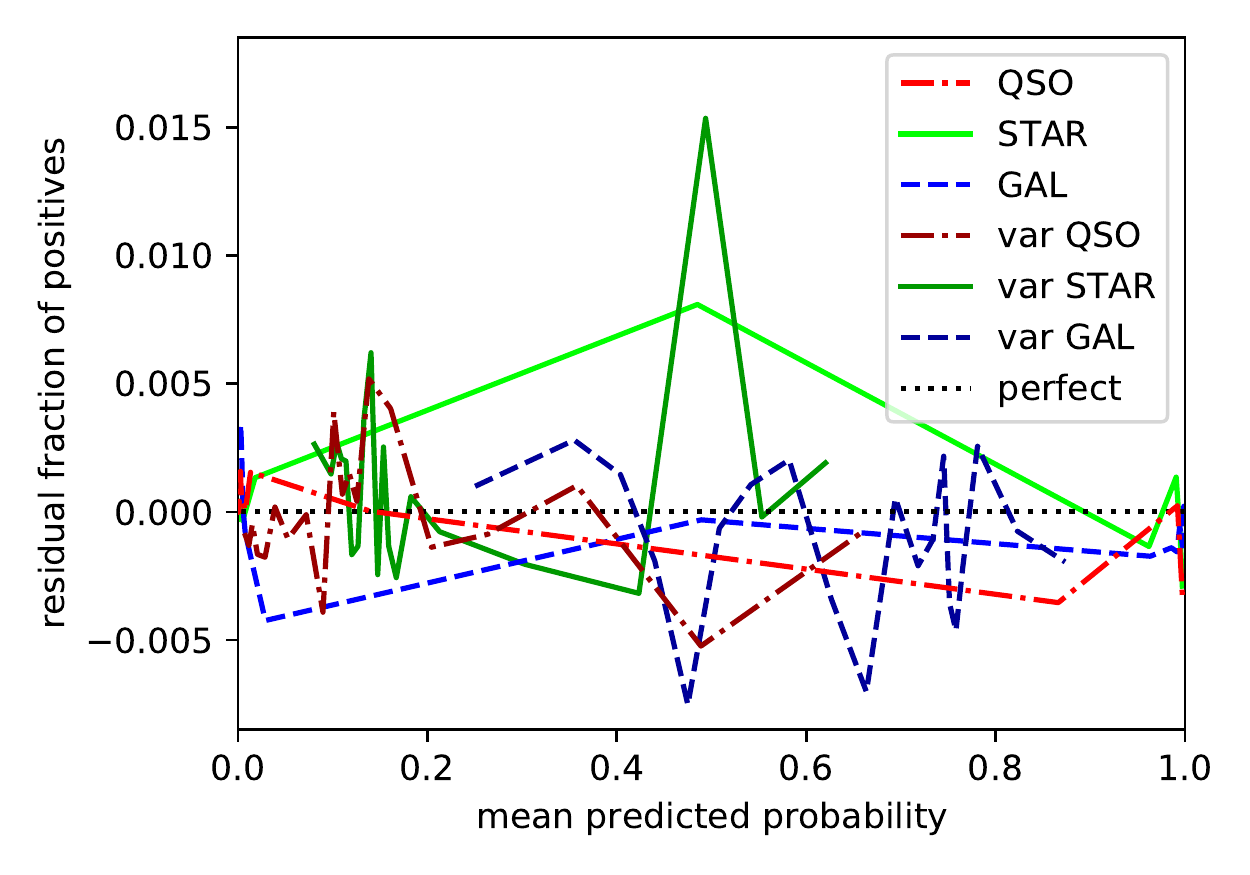}}
	\caption{Calibration plot for the model using all features and all data. As the predictions are very well calibrated, the y axis shows the residuals after subtracting fractions of a perfect model. With only variability, the model is less confident so the predicted probabilities are closer to 50~\% than for models including colours. } 
    \label{fig:calcurve}
\end{figure}

Fig. \ref{fig:calcurve} shows residuals of the calibration curves for each class for the models using all features or only variability on the full dataset with SDSS labels (models AllA and VarA in Sect. \ref{sec:models}). The probabilities of belonging to each class sum to one for every object, as the classes are mutually exclusive. Well calibrated classifiers result in probabilities close to the dotted line where a predicted probability of e.g. 60~\% of being a quasar means that 60~\% of sources with that probability prediction are actually labeled quasars by SDSS.  The bins contain 5 \% of the samples each and are not of equal width. 
The largest residual shows predicted stellar probabilities of $\sim$50~\% being just 1.5 percentage points too low.
As the models are well-calibrated, we do not apply further calibration. The other models predict probabilities of similar quality to AllA and VarA.

For model AllA, the predicted probabilities of being a quasar are in the range 0.00003$-$99.994~\% (over both the train, val and test set). They are 0.0007$-$99.987~\% for stars 
and $0.007-99.9993$~\% for galaxies
. With just variability, the VarA model is not as confident, predicting quasar probabilities of 3$-$80~\%, 
5$-$90~\% 
for stars and 13$-$90~\% 
for galaxies.

\subsection{Predicted parameter distribution} \label{sec:distr}

\begin{figure}[]
	\centering
    \resizebox{\hsize}{!}{\includegraphics{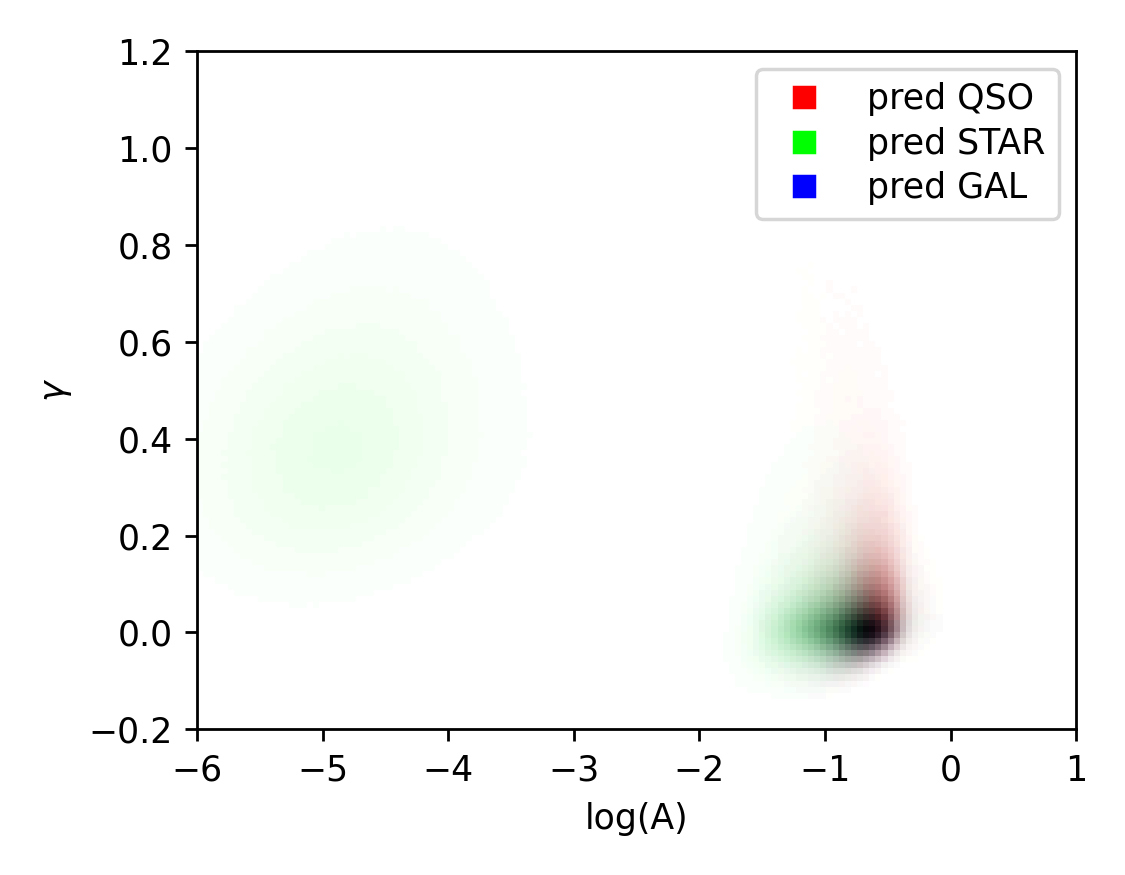}}
    \resizebox{\hsize}{!}{\includegraphics{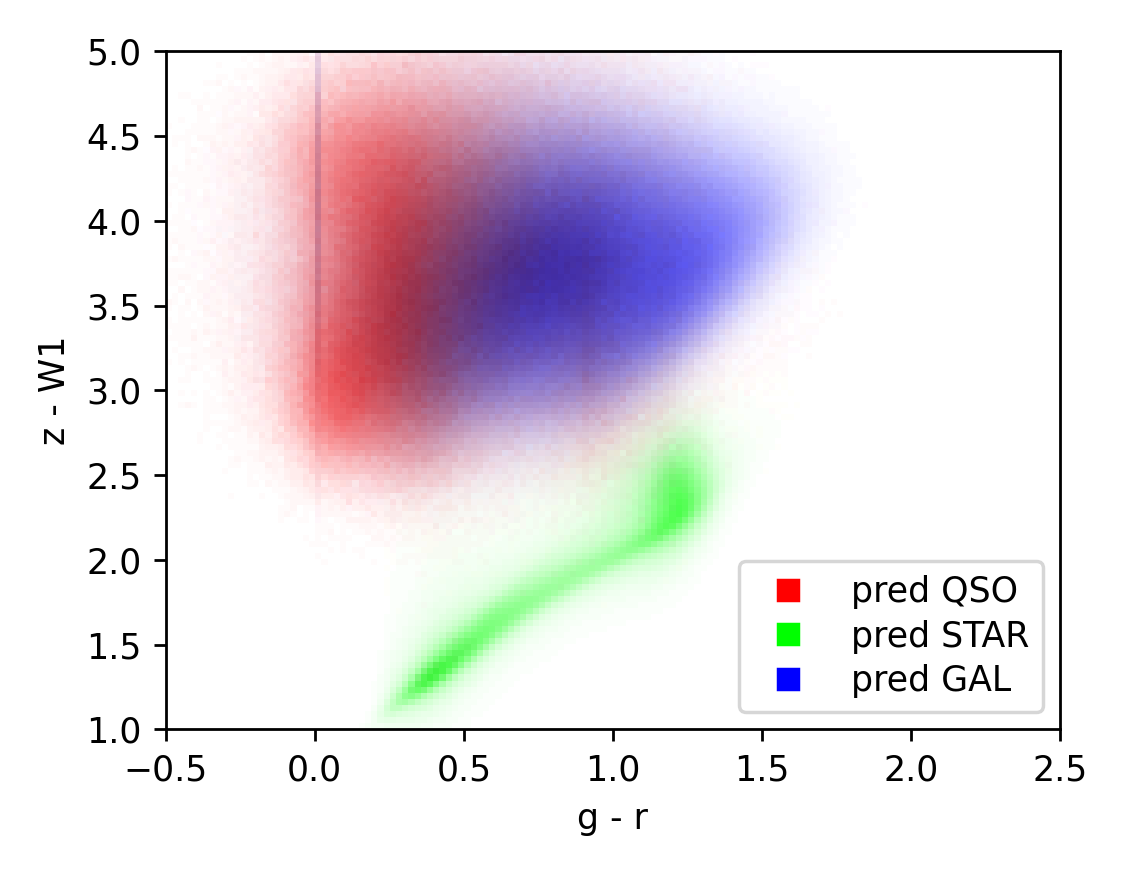}}
    \resizebox{\hsize}{!}{\includegraphics{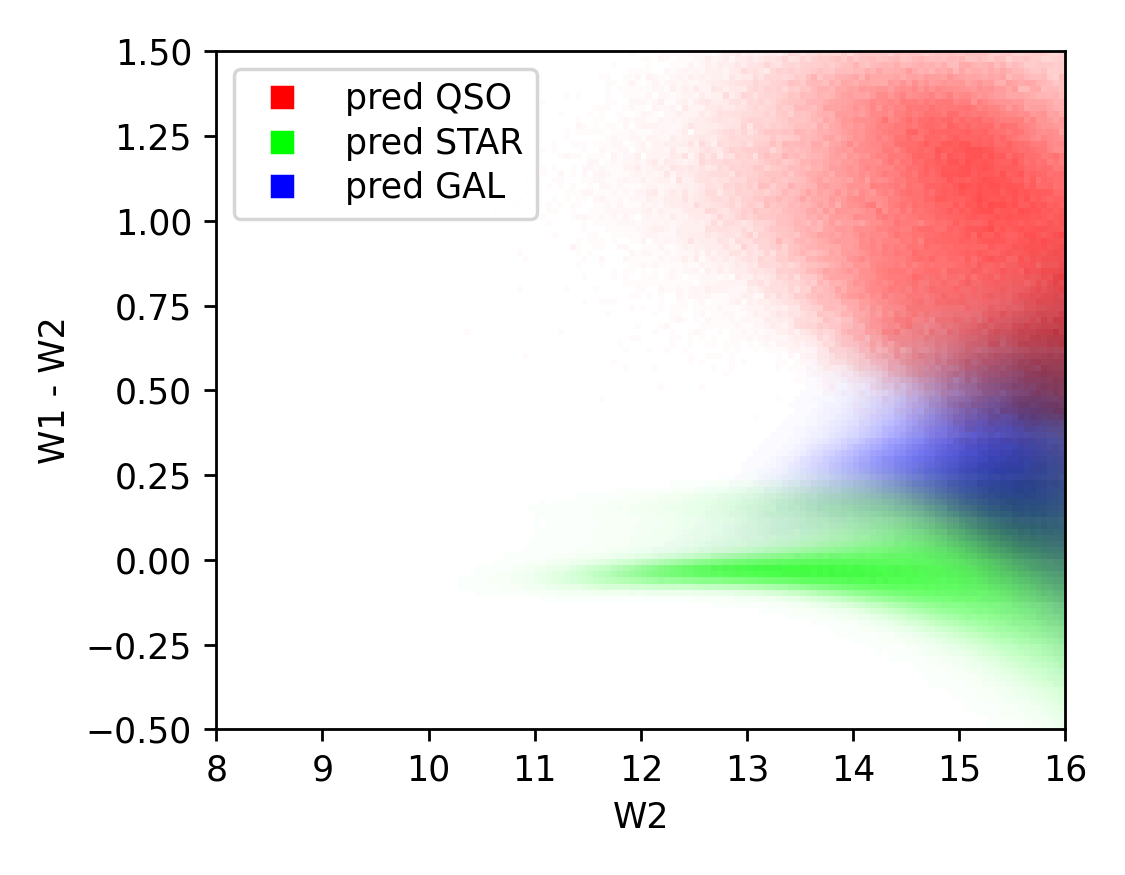}}
	\caption{Predicted classes on all samples using all features, projected in three different parts of the input feature space. With the variability parameters $A$ and $\gamma$ (top) an area at $\log_{10}(A)\sim-0.8$ and $\gamma\sim0$ has high degeneracy, while especially stars and quasars are easier to distinguish in other areas. In $g-r$ vs $z-W1$ (middle) and $W2$ vs. $W1-W2$ (bottom), the model shows greater separation of the classes. These plots include both training, validation and test data.} 
    \label{fig:pred} 
\end{figure} 

In Fig. \ref{fig:pred}, the final predictions of AllA are colour-coded with saturation based on the relative density of each class in the shown parameter space. The plots include the full dataset of 70\,920\,904 samples. The classifier of Sect. \ref{sec:algorithm} selects sources using positions in parameter space in a manner similar to that of paper I -- but optimised automatically to include even the samples with parameters that are not typical of a single class which are therefore difficult to classify (found in grey areas of Fig. \ref{fig:pred}). 

\subsubsection{Variability classification} \label{sec:distr}

\begin{figure}[]
	\centering
    \resizebox{\hsize}{!}{\includegraphics{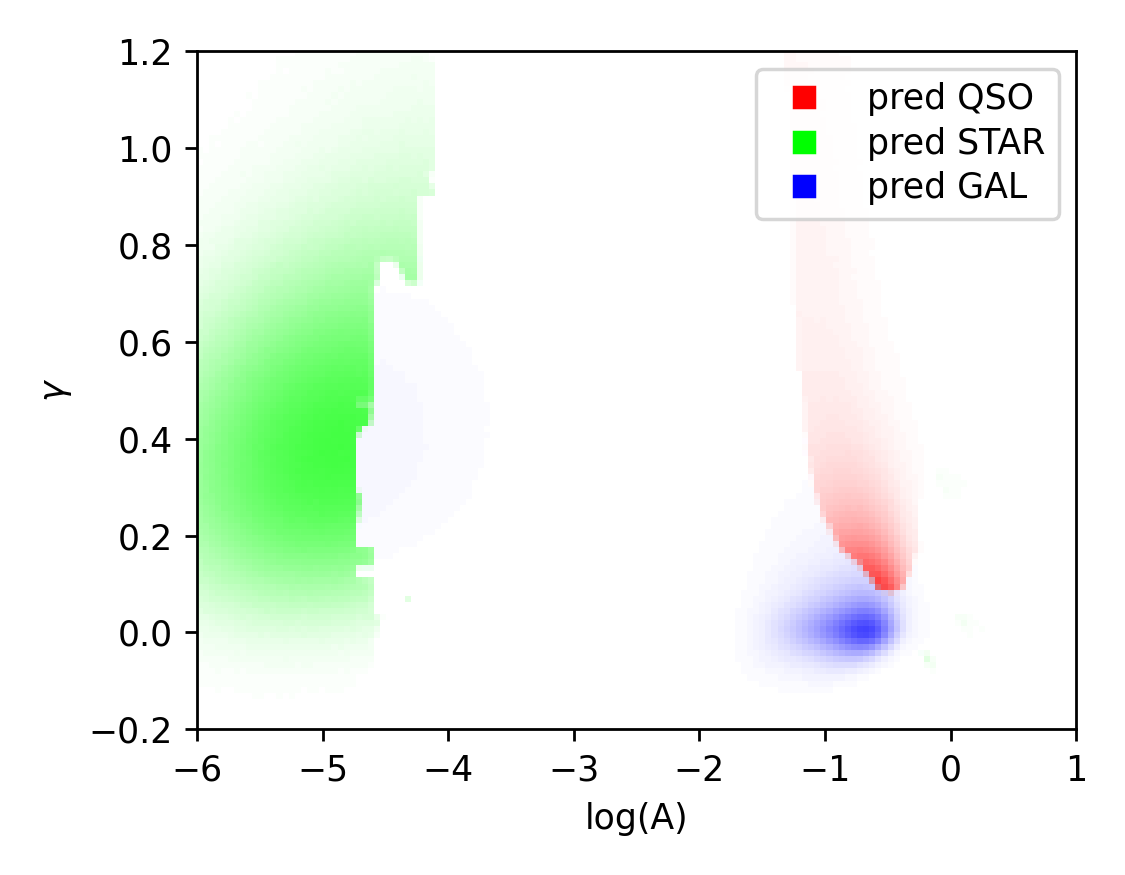}}
	\caption{The model on all samples using only variability predicts classes as illustrated by the diagram above. The model only uses $A$ and $\gamma$ plus combinations of them described in Sect. \ref{sec:preproc}, so the diagram illustrates how the full input feature space maps to model outputs. In most of the feature space, the model predicts the objects to be galaxies, which is also the most frequent class of the dataset.} 
    \label{fig:varpred} 
\end{figure}

Classifying with just the variability parameters $A$ and $\gamma$ (and features combining them) allows us to plot the complete feature space of a model. Fig. \ref{fig:varpred} shows $\log(A)$ vs. $\gamma$ and how VarA separates the classes. Most of the feature space is predicted to contain galaxies. The regions assigned to each class can be compared to those of paper I -- but now, they must cover the entire feature space. The regions are still relatively simple due to regularisation and they approximately match the selection regions of paper I apart from the higher prioritisation of galaxies in sparsely populated regions ($-4.6 < \log (A) < -2$ and $\log (A) > 0$). Areas of low relative frequency of predicted galaxies touch areas of high relative frequencies of predicted stars and quasars, due to the higher total frequency of galaxies in the labeled dataset. 

\subsection{Output table} \label{sec:cat}
Information on outliers, time spans etc. are saved in the final output table as described in Table \ref{tab:output}, along with the resulting predictions and probabilities of each class. The results are produced by model AllA and are published as VILLAIN-Cat in CDS.  
We also select nonvariable sources from two criteria:
\begin{itemize} 
    \item $A$ close to 0: $A+3\sigma_{A,+}<0.01$, and 
    \item $A$ consistent with 0: $A/\sigma_{A,-}<3$.
\end{itemize}

\begin{table*}[]
  \centering
  \caption{Catalogue parameters.}
  \begin{tabular}{l l}
      \hline\hline
      Parameter & Description \\
      \hline
      PTF\_RA & Mean J2000 RA in PTF [deg]\\ 
      PTF\_Dec & Mean J2000 Dec in PTF [deg]\\
      PTF\_ID & Object ID in PTF \\
      \hline
      A & Structure function amplitude of variations on time scales of one year [mag]\\
      dA\_m & Lower error on $A$ [mag]\\
      dA\_p & Upper error on $A$ [mag]\\
      gamma & Power law index \\ 
      dgamma\_m & Lower error on $\gamma$\\
      dgamma\_p & Upper error on $\gamma$\\
      \hline
      Nepochs & Number of $R$ band epochs in PTF after outlier removal\\
      t\_span & Time span in the $R$ band from PTF after outlier removal [days] \\
      median\_R & Median $R$ magnitude from PTF [mag]\\
      outlier\_fraction & Fraction of $R$ band detections classified as outliers\\
      max\_magdiff & Maximum magnitude difference from the $R$ band moving median [mag]\\
      mean\_magdiff & Mean magnitude difference from the $R$ band moving median [mag]\\ 
      \hline
      W1 & WISE $W1$ band [mag]\\
      dW1 & Error on $W1$ [mag]\\
      W2 & WISE $W2$ band [mag]\\
      dW2 & Error on $W2$ [mag]\\
      \hline
      g & PS1 $g$ band [mag]\\
      dg & Error on $g$ [mag]\\
      r & PS1 $r$ band [mag]\\
      dr & Error on $r$ [mag]\\
      i & PS1 $i$ band [mag]\\
      di & Error on $i$ [mag]\\
      z & PS1 $z$ band [mag]\\
      dz & Error on $z$ [mag]\\
      y & PS1 $y$ band [mag]\\
      dy & Error on $y$ [mag]\\
      \hline
      SDSS\_RA & J2000 RA to SDSS match [deg]\\
      SDSS\_Dec & J2000 RA to SDSS match [deg]\\
      redshift & Spectroscopic SDSS redshift \\
      redshiftErr & Redshift error\\ 
      SDSS\_class & Spectroscopic classification from SDSS\\
      \hline
      nonvariable & 1 if $A$ is low and consistent with zero (see Sect. \textcolor{blue}{\ref{sec:cat}})\\
      \hline
      P\_QSO & Probability of being a quasar\\
      P\_GAL & Probability of being a galaxy\\
      P\_STAR & Probability of being a star\\
      pred   & Predicted class
  \end{tabular}
  \tablefoot{Parameters of the output catalogue published as VILLAIN-Cat. For SDSS\_class and pred, quasars are labeled 0, stars 1 and galaxies 2. Unknown values are included as NaN.}
  \label{tab:output}
\end{table*}{}
  
\subsection{Feature importance}

We measure the contribution of features by dropping some features in models 3 -- 6 of Sect. \ref{sec:models} and with permutation importance. With permutation importance, each feature is evaluated by randomly permuting its values and measuring the macro averaged F1 loss. This measure has a bias towards correlated features \citep{strobl,nicodemus}, so important but highly correlated features will likely all get low rankings. It is, however, less resource intensive than computing Shapley values \citep{shapley} which can be implemented in ways that minimise the bias \citep{CoShapley,ACV}.  

The top 20 permutation importances of AllA are listed in Table \ref{tab:feature_ranking}. The top  13 features are all engineered, including two based on selections in paper I. 
The highest ranking variability feature is $\gamma-2\log (A)$ in the 16th place.  
The sum of the losses is much smaller than the total F1 score of 0.9639, showing that most information lies in highly correlated features. $R-r$ might be ranked high because it can be used for estimation of stellarity since the image quality of PTF and PS1 is different. 

\begin{table}[]
      \centering
      \caption{Feature importance.}
      
      \begin{tabular}{l l}
          \hline\hline
          Feature & Loss \\
          \hline
          $R-r$ & $0.045$\\
          $z-W1$ & $0.032$\\
          $z-W1-1.25(g-r)$ & $0.029$\\
          $r-i$ & $0.013$\\
          $i-z$ & $0.010$\\
          $R-W2$ & $0.008$\\
          $z-W2$ & $0.007$\\
          $R-i$ & $0.004$\\
          $R-g$ & $0.004$ \\
          $y-W1$ & $0.002$ \\
          $W1-W2-0.017W2$ & $0.002$ \\
          $r-z$ & $0.002$ \\
          $y-W2$ & $0.002$ \\
          $W2$ & $0.001$ \\
          $i$ & $0.001$ \\
          $\gamma-2\log (A)$ & $0.001$ \\
          $\gamma+0.5\log (A)$ & $0.001$ \\
          $\gamma$ & $0.001$ \\
          $g$ & $0.001$ \\
          $z$ & $0.001$ 
      \end{tabular}
      
      \label{tab:feature_ranking}
      \raggedright
        \tablefoot{Permutation importance of the top 20 features. Importance is evaluated with the model using all features and all data, on the test set. All uncertainties are smaller than 0.0005. The losses are highly biased due to feature correlations, but indicate that engineered features are important. $R-r$ could work as a measure of stellarity.  
        }
  \end{table}{}

\section{Discussion and conclusions} \label{sec:discussion}

Using PTF variability, magnitudes from PTF, WISE and PS1 and their combinations, we have created a model (AllA in Sect. \ref{sec:models}) that identifies 1\,330\,412 quasar, 48\,618\,737 star and 20\,971\,755 galaxy candidates with a macro averaged F1 score of 0.9639 on spectroscopically classified sources from SDSS. The model detects 92.49~\% of SDSS quasars and has an overlap with SDSS quasars of 95.64~\%. We are able to assess the quality of the classifications using predicted probabilities, which are highly accurate, as described in Sect. \ref{sec:calibration}, and allow for accurate discrimination of true and false positives, as described in Sect. \ref{sec:roc}.

\subsection{Model comparison}

According to Table \ref{tab:scores}, performance is slightly improved by including variability, especially for quasars. The quasar F1 score improves from 0.9570 to 0.9580 on the matched samples (MagM vs. AllM). The importance of colour is much greater, shown by the lower performance of variability-only models, giving a quasar F1 score of 0.4462 on the matched samples (VarM). This is still better than the baseline for random guessing of 0.33.
Using all features and data (AllA), we are able to give confident predictions of up to 99.994~\% as quasar probability -- which is also much better than the 80~\% with only variability (VarA). By including variability in models, however, we identify more candidates, as more samples can be used. Fig. \ref{fig:varpred} shows how classification is performed with only variability. 

For objects with variability parameters, one could always include a feature with the average magnitude in the lightcurve to improve performance with e.g. a variability + $R$ model. This would likely identify more bright objects as stars.

This study is on PTF sources with measured variability. If variability was not required, including colours would also enable the classification of more sources, i.e. sources without variability information. Although the presented models are classifying all samples, we still see higher purities, completenesses and clearer visible separation of classes in Fig. \ref{fig:pred} than with the limited manual selections in Figs. 2--3 of paper I.

\subsection{SDSS comparison}

The candidate parameter distributions of Fig. \ref{fig:pred} are close to the distributions in SDSS labeled objects in Fig. 2 of paper I, but with different total distributions as the sources only have to be found in PTF (Fig. 1 in paper I).  
We find relatively more galaxies at $W1-W2>0.5$ and $W2>15$ and a more unified cluster of galaxies in $g-r$ vs. $z-W1$ compared to SDSS. 
The small structures at $\gamma>0.5$ and $\log A>-0.2$ in Fig. 1 of paper I are mostly classified as stars. We judge the structure at $g-r=0$ to be an artefact.

\subsection{Biases} \label{sec:biases}

Each survey has selection effects, affecting ratios of the object types and their parameters. 
Performance is better for sources matched in PS1 and WISE,  
indicating that matched sources are generally easier to classify. 
Variability might perform differently compared to colours, if we change the outlier removal and constraints on e.g. lightcurve length by paper I.

We also have a bias from the manual exploration of the same dataset in paper I, including the test set of this paper. The models are therefore not created entirely independently of the test set, but they have only been tested on it once. 

Each PTF lightcurve is constructed in paper I by grouping data points with the same PTF ID. All lightcurves have different PTF ID's, so all samples should correspond to independent astrophysical objects. However, some PTF objects have near-identical coordinates. For these ID's, PTF lightcurves are therefore correlated, and the objects are likely matched to the same object in WISE and PS1. 
This gives a small bias, since correlated feature values appear across the training and test set. For labeled data, all SDSS matches are unique, but 0.02~\% (268) of the sources have a duplicate match in WISE and 0.004~\% (70) have a duplicate in PS1. 22~\% (25\,329\,144) of PTF objects have a neighbour within two arcseconds, but only 0.01~\% (158) within the labeled set. 

\subsection{Perspectives}
The catalogue described in Sect. \ref{sec:cat} includes predictions and probabilities of 70\,920\,904 PTF objects. It greatly expands on the set of 4\,618\,756 DR17 SDSS spectroscopic 
classifications out of which the labels are from 1\,747\,471. The catalogue includes confident quasar candidates (high P\_QSO) and standard star candidates (flagged nonvariable) for use in future research. Subsets could be interesting to observe and analyse further, such as quasar candidates or variable galaxies. Using the included probabilities, one can extract a set of e.g. quasar candidates of a chosen minimum P\_QSO for a preferred trade-off between completeness and purity.

Monochromatic variability is not enough for confident classification of most sources into the three macro-classes. However, it does add information within each class and could be used for selection of rare subtypes or distinction of e.g. type I and type II AGN \citep{cicco2022}. It is also useful in cases where colours are not available. For simple variability selection without machine learning, we suggest selecting by regions similar to those of Fig. \ref{fig:varpred}.

For machine learning tasks on similar data, we suggest datasets of at least $\sim$100\,000 labeled samples for an expected macro averaged F1 score of 0.9610 for new data. The required survey size for a given score can be estimated from Fig. \ref{fig:N_labeled}. For small datasets, we suggest adjusting class ratios by oversampling (copying or generating synthetic samples) 
or more advanced techniques combining oversampling and undersampling (removing samples) such as SMOTETomek or SMOTEENN \citep{SMOTETomek} on the training set (not on the test set). We also suggest combining the training and validation sets before testing on small sets and using cross-validation. Stratification can ensure that even small folds are representative of all classes. \texttt{HistGradientBoostingClassifier} is only faster than \texttt{GradientBoostingClassifier} from scikit-learn for datasets of $\geq$10\,000 samples, so alternative algorithms may be considered for speed. For large numbers of classes, both algorithms are inefficient since they create a tree for each class during each boosting iteration. If a large unlabeled set is available, semi-supervised learning can be applied to learn from it. In that case, the unlabeled samples can all be added to the training set. 

In areas of parameter space with few SDSS labels, performance can be improved using active learning \citep{activelearning}. By identifying where in feature space new labels would be most useful, and expanding the labeled set accordingly, fewer labeled samples are needed. 
This is especially relevant in areas with higher relative population densities in PTF than SDSS or, in the future, in the LSST. 
Performance would of course also be improved by adding additional information or well-designed feature engineering such as more bands, photometric redshifts, proper motions, stellarity etc. The joint survey processing of LSST, \textit{Euclid} \citep{euclid} and the \textit{Nancy Grace Roman Space Telescope} \citep{roman} will include deep, multi-band information in the optical and near-infrared, which can be used similarly to the WISE and PS1 bands in this project \citep{JSP}. Another approach to automatic classification with variability is using time-series layers in neural networks. More information can be captured by the model by directly using the full lightcurves instead of, or in addition to, manually selected summarising features like $A$ and $\gamma$ \citep{timeNNastro}.

We used a histogram based gradient boosting classification tree, which is fast, performs well, learns from missing values, produces probabilities, detects nonlinear patterns and is easy to implement with scikit-learn. Feature engineering further improves performance. No other astronomical papers in the SAO/NASA Astrophysics Data System (ADS)\footnote{https://ui.adsabs.harvard.edu} mention this implementation to date, but we recommend further use in astronomy as an alternative to XGBoost and LightGBM. With models like the ones of this paper, future sources can quickly and automatically be classified.

\begin{acknowledgements} 

    This work was supported by a Villum Investigator grant (project number 16599).
    SHB was also supported by a grant from the Danish National Research Foundation.
    AA was also supported by a Villum Experiment grant (project number 36225). 

    This publication makes use of data products from the Wide-field Infrared Survey Explorer, which is a joint project of the University of California, Los Angeles, and the Jet Propulsion Laboratory/California Institute of Technology, funded by the National Aeronautics and Space Administration. 

    The Pan-STARRS1 Surveys (PS1) and the PS1 public science archive have been made possible through contributions by the Institute for Astronomy, the University of Hawaii, the Pan-STARRS Project Office, the Max-Planck Society and its participating institutes, the Max Planck Institute for Astronomy, Heidelberg and the Max Planck Institute for Extraterrestrial Physics, Garching, The Johns Hopkins University, Durham University, the University of Edinburgh, the Queen's University Belfast, the Harvard-Smithsonian Center for Astrophysics, the Las Cumbres Observatory Global Telescope Network Incorporated, the National Central University of Taiwan, the Space Telescope Science Institute, the National Aeronautics and Space Administration under Grant No. NNX08AR22G issued through the Planetary Science Division of the NASA Science Mission Directorate, the National Science Foundation Grant No. AST-1238877, the University of Maryland, Eotvos Lorand University (ELTE), the Los Alamos National Laboratory, and the Gordon and Betty Moore Foundation. 
  
    Funding for the Sloan Digital Sky 
    Survey IV has been provided by the 
    Alfred P. Sloan Foundation, the U.S. 
    Department of Energy Office of 
    Science, and the Participating 
    Institutions. 

    SDSS-IV acknowledges support and 
    resources from the Center for High 
    Performance Computing  at the 
    University of Utah. The SDSS 
    website is www.sdss.org. 
    
    SDSS-IV is managed by the 
    Astrophysical Research Consortium 
    for the Participating Institutions 
    of the SDSS Collaboration including 
    the Brazilian Participation Group, 
    the Carnegie Institution for Science, 
    Carnegie Mellon University, Center for 
    Astrophysics | Harvard \& 
    Smithsonian, the Chilean Participation 
    Group, the French Participation Group, 
    Instituto de Astrof\'isica de 
    Canarias, The Johns Hopkins 
    University, Kavli Institute for the 
    Physics and Mathematics of the 
    Universe (IPMU) / University of 
    Tokyo, the Korean Participation Group, 
    Lawrence Berkeley National Laboratory, 
    Leibniz Institut f\"ur Astrophysik 
    Potsdam (AIP),  Max-Planck-Institut 
    f\"ur Astronomie (MPIA Heidelberg), 
    Max-Planck-Institut f\"ur 
    Astrophysik (MPA Garching), 
    Max-Planck-Institut f\"ur 
    Extraterrestrische Physik (MPE), 
    National Astronomical Observatories of 
    China, New Mexico State University, 
    New York University, University of 
    Notre Dame, Observat\'ario 
    Nacional / MCTI, The Ohio State 
    University, Pennsylvania State 
    University, Shanghai 
    Astronomical Observatory, United 
    Kingdom Participation Group, 
    Universidad Nacional Aut\'onoma 
    de M\'exico, University of Arizona, 
    University of Colorado Boulder, 
    University of Oxford, University of 
    Portsmouth, University of Utah, 
    University of Virginia, University 
    of Washington, University of 
    Wisconsin, Vanderbilt University, 
    and Yale University. 
    
    This research made use of the cross-match service provided by CDS, Strasbourg.
    
    This research made use of Astropy,\footnote{http://www.astropy.org} a community-developed core Python package for Astronomy (\citeauthor{Astropy1}, \citeyear{Astropy1}; \citeauthor{Astropy2}, \citeyear{Astropy2}). 
    
    This research has made use of the NASA/IPAC Infrared Science Archive, which is funded by the National Aeronautics and Space Administration and operated by the California Institute of Technology. 
    
    This research makes use of the SciServer science platform (www.sciserver.org). 

    SciServer is a collaborative research environment for large-scale data-driven science. It is being developed at, and administered by, the Institute for Data Intensive Engineering and Science at Johns Hopkins University. SciServer is funded by the National Science Foundation through the Data Infrastructure Building Blocks (DIBBs) program and others, as well as by the Alfred P. Sloan Foundation and the Gordon and Betty Moore Foundation. 
    
    This research has made use of the VizieR catalogue access tool, CDS, Strasbourg, France (DOI : 10.26093/cds/vizier). The original description of the VizieR service was published in 2000, A\&AS 143, 23. 
     
    This research has made use of NASA’s Astrophysics Data System.
\end{acknowledgements}


\bibliographystyle{aa} 
\bibliography{bibliography.bib}

\end{document}